\documentstyle[12pt]{article}
\newcommand{\be}{\begin{equation}}
\newcommand{\ee}{\end{equation}}
\newcommand{\ba}{\begin{eqnarray}}
\newcommand{\ea}{\end{eqnarray}}
\newcommand{\n}{\nonumber\\}

\newcommand{\br}{\mbox{{\bf R}}}

\begin{document}
\begin{flushright}
{La Plata Th/02-04\\November, 2002}
\end{flushright}


\begin{center}

{\Large\bf Supertubes and special holonomy}

\bigskip

{\it \large Nicol\'{a}s E. Grandi\footnote{
{\sf grandi@fisica.unlp.edu.ar; Fellow CONICET, Argentina.}
} and \large Adri\'{a}n R. Lugo\footnote{
{\sf lugo@fisica.unlp.edu.ar; Partially supported by CONICET, Argentina.  }
}}

\bigskip

{\it Departamento de F\'\i sica, Facultad de Ciencias Exactas \\
Universidad Nacional de La Plata\\ C.C. 67, (1900) La Plata,
Argentina}
\bigskip

\end{center}
%
%
%
%
\begin{abstract}
We obtain a $\frac 14$-supersymmetric $6$-brane solution of IIA Supergravity by T-dualizing the supertube recently
found. The resulting $C_{\it 1}$ electric charge is related to the original $D0$-brane charge. The uplifted
solution to eleven dimensions results to be a purely geometrical configuration, which can be interpreted as a
bound state of a Taub-NUT space and a pp-wave. Being the non trivial part of the metric pseudo-Riemannian, the
resulting reduced holonomy group is non-compact and locally isomorphic to a semidirect product of an Abelian four
dimensional group and $SU(2)$.
\end{abstract}

\section{Introduction}

\bigskip

Tubular configurations of $D2$-branes preserving some supersymmetries
became a subject of interest in recent years \cite{Mateos:2001qs}-\cite{Mateos:2002yf}.
The supertube configuration of circular cross-section was introduced in
\cite{Mateos:2001qs} from the world-volume point of view, and its corresponding
supergravity solution was presented in \cite{Emparan:2001ux}.
In \cite{Mateos:2001pi} the generalization to arbitrary cross-section was found.

The supertube has zero $D2$ charge, which agrees with its interpretation as the ``blown-up'' configuration arising
when $D0$ charge is disolved in a bundle of $F1$ superstrings. Being $\frac 14$-supersymmetric, the blown-up
configuration is self supported against collapse.

The T-dual configuration along the longitudinal axis of the supertube has been presented
in \cite{Cho:2001ys}; from the author's analysis of the boundary conditions arising on a superstring
that ends on the supertube the picture of a ``super $D$-helix'' emerged.
The ST-dual of the super D-helix was studied in \cite{Mateos:2002yf} and
identified with the type IIA supercurve, i.e. a string with an arbitrary transverse
displacement traveling at the speed of light.

In the present paper a $D6$-brane configuration is obtained by T-dualizing the supertube.
Somewhat surprisingly, it can be uplifted to a purely geometrical eleven dimensional
solution, that in view of the supersymmetries preserved by the original supertube has
a special holonomy group.

\bigskip

\section{The supertube solution}

The supergravity solution sourced by $N$ coincident supertubes with arbitrary
cross section is given by \cite{Mateos:2001pi}
\ba
ds^2_{\it 10} &=& - V^{-1/2} U^{-1}  \, ( dt - A)^2 +
V^{1/2}(U^{-1}  \, dz^2 + d\vec{y} \cdot d\vec{y}+ d\vec{x} \cdot d\vec{x} )\, \cr
B_{\it 2} &=& - U^{-1} \, (dt - A) \wedge dz + dt \wedge dz\, \cr
C_{\it 1} &=&  V^{-1} \, (dt - A) - dt \, \cr
C_{\it 3} &=& - U^{-1} dt\wedge dz \wedge A \, \cr
e^\phi &=& U^{-1/2} V^{3/4}
\label{solution}
\ea
where the coordinates of ten dimensional spacetime are $(t,z,\vec y, \vec x)$.
The supertube we will consider has the topology ${\cal C}\times {\br}$ where the $\br$
direction is identified with the $z$ variable and ${\cal C}$ is an arbitrary
curve embedded in the $\vec y= (y_1,y_2,y_3)$ hyperplane; $\vec x \in \br^5$ are
coordinates transverse to the supertube.

The harmonic functions $U(\vec x, \vec y)$, $V(\vec x, \vec y)$
and the 1-form $A= \vec A(\vec x,\vec y).d\vec {y}$ are given by:
%
%
\be
\left.
\begin{array}{r}
U(\vec x,\vec y)-1\cr
V(\vec x,\vec y)-1\cr
\vec A(\vec x,\vec y)
\end{array}
\right\}
= \frac 1{6\Omega_7}\int_{\cal C} d\sigma\; \frac 1{R(\vec x,\vec y;\sigma)^6}
\left\{
\begin{array}{l}
|\pi(\sigma)|\cr
|B(\sigma)|\cr
N{\vec y}'(\sigma)
\end{array}
\right.
\label{funciones}
\ee
where we have defined $R(\vec x,\vec y;\sigma)= \sqrt{(\vec y -\vec y(\sigma))\cdot(\vec y- \vec y(\sigma))+ \vec x\cdot\vec x}$,
and the cross-section ${\cal C}$ of the supertube is specified by the functions $\vec{y} = \vec{y}(\sigma)$, the
$F1$-string and $D0$-brane charge densities are denoted $\pi(\sigma)$ and $B(\sigma)$ respectively. Here $\Omega_{\it q}$
is the volume of a $q$-sphere of unit radius.

The gauge-invariant field-strengths derived form the Neveu-Schwarz $B_{\it 2}$ and
Ramond-Ramond $C_{\it p}$ forms are given by
\ba
H_{\it 3}=dB_{\it 2} \; \; \;, \;\;\;
G_{\it 2}=dC_{\it 1} \; \; \;, \;\;\;
G_{\it 4}=dC_{\it 3} + H_{\it 3} \wedge C_{\it 1}
\ea
For the solution (\ref{solution}) they read
\ba
H_{\it 3} &=& U^{-2} dU\wedge (dt-A)\wedge dz + U^{-1}dA\wedge dz\,\cr
G_{\it 2} &=& -V^{-2}dV\wedge(dt-A) - V^{-1}dA\,\cr
G_{\it 4} &=& -U^{-1} V^{-1} (dt -A) \wedge dz \wedge dA
\label{g4}
\ea

The Hodge duals of the above RR-curvatures are
\ba
G_{\it 6} &\equiv& *_{\it 10}G_{\it 4}\equiv dC_{\it 5} + H_{\it 3}\wedge C_{\it 3}=  *_{\it 8} dA
\cr
G_{\it 8} &\equiv& *_{\it 10}G_{\it 2}\equiv -dC_{\it 7} - H_{\it 3}\wedge C_{\it 5}=   \left(-U^{-1}  G_{\it 6}\wedge(dt-A) - *_{\it 8} dV\right)\wedge dz
\ea
where $*_{\it 8}$ stands for the eight-dimensional Hodge dual with respect to the Euclidean $(\vec y,\vec x)$
space and the dual potentials, defined according to the equations of motion\footnote{We follow conventions of
\cite{Johnson:2000ch}.}, are given by
\ba
C_{\it 5}&=& *_{\it 8}T_{\it 3}\,\cr
C_{\it 7}&=& *_{\it 8}T_{\it 2}\wedge dz+U^{-1}C_{\it 5}\wedge (dt-A)\wedge dz
\label{dualpot}
\ea
where we have introduced the two new forms $T_{\it 2}$ and $T_{\it 3}$ that must satisfy
the constraints
\ba
d *_{\it 8}T_{\it 3}=  *_{\it 8}dA
\;\;\; , \;\;\;
d *_{\it 8} T_{\it 2} = *_{\it 8} dV
\label{constraints}
\ea

\section{T-duality}

Let $\;(x^\mu , y)\;$ be coordinates in which a $y$-translation is a symmetry of the
whole background (metric $G_{MN}$, antisymmetric tensor $B_{MN}$, dilaton $\phi$,
RR fields $C_{\it p}$) and $(x^\mu)$
coordinates in nine space-time dimensions.
Then to perform a T-duality in $y$-direction we use the following form of the duality
rules; on NS-NS fields
\ba
\tilde {ds}_{\it 10}^2 &=& ds_{\it 9}^2 + G_{yy}{}^{-1}\;(b + d{y})^2 -
G_{yy}\;g^2\cr
\tilde B_{\it 2} &=& B_{\it 2}|_{\it 9} + g\wedge (b + dy)\cr
e^{{\tilde\phi}} &=& G_{yy}{}^{-{1}/{2}}\; e^{{\phi}}\label{TdualityNeveu}
\ea
where
$\; g\equiv{G_{yy}}^{-1} {G_{\mu y}}\,dx^\mu
\;$, $\; b\equiv  B_{\mu y}\,dx^\mu\;$ and $B_{\it 2}|_{\it 9}=B_{\mu\nu}\,dx^\mu \wedge dx^\nu$,
and on RR-fields
\ba
\tilde{C}_{\it n} &=& (\partial_y\,;C_{\it n+1}) + (\partial_y\,;C_{\it n-1})\wedge (b+dy)\wedge g
+ (\partial_y\,; C_{\it n-1}\wedge dy )\wedge b
+C_{\it n-1}\wedge dy
\ea
where $\;(u\, ; t_{\it n+1})_{\mu_1\dots\mu_n}\equiv u^\mu\; t_{\mu_1\dots\mu_n\mu}\;$.

We will use these rules to T-dualize the solution (\ref{solution}) in the directions $(z,\vec x)$.
After the first T-duality in the $z$ direction, the resulting non-zero
fields are
\ba
ds_{\it 10}^2&=&  V^{-1/2}\left((U-1) {dz^+}^2
 - dz^+(dz^--2A)\right)
 +V^{1/2}(d\vec y\cdot d\vec y+ d\vec x \cdot d\vec x)\cr
C_{\it 2}&=&\frac 12 \left(V^{-1}(dz^- -2A)- dz^- \right)\wedge dz^+\cr
C_{\it 6}&=& C_{\it 5}\wedge dz^++*_{\it 8}T_{\it 2}\cr
e^\phi &=& V^{1/2}
\label{zdualized}
\ea
where $z^{\pm}= t \pm z$, $C_{\it 5}$ and $T_{\it 2}$ are given in (\ref{dualpot}) and (\ref{constraints}) respectively.
This supergravity solution corresponds to the super D-helix first studied in \cite{Cho:2001ys},
the ST-dual of which is the supercurve of \cite{Mateos:2002yf}.
Here we see that the $B_{\it 2}$ form vanishes. This is due to the fact that the $F$-strings
extending in the $z$ direction are transformed into light-like momentum modes in the $z$ direction
after the T-dualization.

In order to perform a T-duality in a given direction, the fields have to be independent of
the corresponding variable. This is achieved by {\it delocalizing} the solution, i.e. forming
an array of supertubes in the direction to be T-dualized and taking the limit in which the separating distance goes
to zero and the charge densities remain constant. For each one the $\vec x$ directions, the net effect of
these operations in (\ref{funciones}) is to decrease the power of $R(\sigma)$ in
one, eliminating the corresponding dependence.
The $(q-1)\Omega_q$ denominator in the solutions goes to $(q-2)\Omega_{q-1}$
on each step.
After delocalizing the supertube in the five $\vec x$ directions we obtain
%

\be
\left.
\begin{array}{r}
U(\vec y)-1\cr
V(\vec y)-1\cr
\vec A(\vec y)
\end{array}
\right\}
= \frac 1{4\pi}\int_{\cal C} d\sigma\; \frac 1{|\vec y- y(\sigma)|}
\left\{
\begin{array}{l}
|\pi(\sigma)|\cr
|B(\sigma)|\cr
N{\vec y}'(\sigma)
\end{array}
\right.
\label{funcionesdelocalizadas}
\ee

It will be convenient in what follows to rewrite the $T_{\it 2}$ and $T_{\it 3}$
forms in terms of their 3-dimensional Hodge duals w.r.t. the $\vec y$ variables, $T_{\it 1} = *_{\it 3}T_{\it 2}
=\vec T\cdot  {d \vec y}$ and $T= *_{\it 3}T_{\it 3}$, then
the constraints (\ref{constraints}) are written as
\ba
\vec\nabla T = \vec\nabla\times
\vec A\;\;\; ,\;\;\;
\vec\nabla\times\vec T = \vec\nabla V\label{constraintsbis}
\ea
Next we will T-dualize (\ref{zdualized}) in the five $\vec x$ directions, which are transverse to the supertube.
We obtain
\ba
ds_{\it 10}^2 &=& V^{-1/2}\left((U-1) {dz^+}^2 - dz^+ (dz^--2 A)
+d\vec x \cdot d\vec x \right) + V^{1/2} d\vec y \cdot d\vec y\n
C_{\it 1} &=& T_{\it 1} - T\; dz^+\n
C_{\it 7} &=& \frac 12\left( V^{-1}(dz^--2A)-dz^-  \right)\wedge dz^+\wedge dx_1 \wedge \dots\wedge dx_5\cr
e^\phi &=& V^{-3/4}\,.
\label{final}
\ea
Let us remark here that the NS 2-form $B_{\it 2}$ and the RR 3-form $C_{\it 3}$ vanish,
which implies that this solution can be uplifted to a
purely geometrical eleven dimensional configuration, as we will see in the next section.

\section{Eleven dimensional metric and reduced holonomy}

\subsection*{{Uplifting to eleven dimensions}}
Let us carry out the uplifting to $D=11$ of the $D6$-brane like solution of
$D=10$ SUGRA IIA just found, according to the standard recipe
\ba
ds_{\it 11}^2&=& e^{-\frac{2}{3}\phi}\;ds_{\it 10}^2 + e^{\frac{4}{3}\phi}\;
\left(d\psi + C_{\it 1}\right)^2\cr
A_{\it 3} &=& C_{\it 3} + B_{\it 2}\wedge d\psi
\ea
where $\psi$ is the eleventh dimension.
Being $ C_{\it 3} = B_{\it 2} =0$ we get $A_{\it 3}=0$, i.e.  the solution is a pure geometric one. It reads
\ba
ds^2_{\it 11} &=& ds^2_{\it 1,5} + d\vec x\cdot d\vec x \cr
ds^2_{\it 1,5} &=& \left(U-1+V^{-1}T^2\right)\,{dz^+}^2 - dz^+\left(
dz^- +\;{\cal A}\right)
+V\; d\vec y\cdot d\vec y + V^{-1}\;\left( d\psi +  T_{\it 1}\right)^2
\label{metrica15}
\ea
where ${\cal A}=-2A+ 2\,V^{-1}T (d\psi + T_{\it 1})$ is the so called Sagnac connection \cite{Gibbons:1999uv}.

It will be convenient in what follows to work in a local basis, so let us introduce the
following elfbein
$\{\omega^A\}$ together with the dual basis $\{e_A\}$ in $T(M)$,  $\;(e_A;\omega^B)=\delta^B_A$,
\ba
&&\!\!\!\!\!\!\!\!\!\!\!\!
\begin{array}{ll}
\omega^0 = U^{-\frac{1}{2}}\; (dt - A)&
e_0 = U^{1/2}(\partial_t-(1-U^{-1})\partial_z + U^{-1}T \partial_\psi)
\cr
\omega^i = V^{\frac{1}{2}}\;dy^i &
e_i  = V^{-1/2}(\partial_i+ A_i(\partial_t-\partial_z)-T_i \partial_\psi)
\cr
\omega^4 = V^{-\frac{1}{2}}\; \left(dx^{10} + C_{\it 1}\right)&
e_4 = V^{1/2}\partial_\psi
\cr
\omega^5 = U^{-\frac{1}{2}}\; \left( U\;dz + (U-1)\;dt + A\right)&
e_5 = U^{-1/2}(\partial_z + T\partial_\psi)
\cr
\omega^a = dx^{a-5}
& e_a = \partial_{x^{a-5}}
\end{array}
\cr
&&
\label{elfbein11}
\ea
where $i=1,2,3$ and $a=6,\dots,10$ and  $C_{\it 1}$ is given in (\ref{final}).
The metric (\ref{metrica15}) is simply
\be
ds^2_{\it 11} = \eta_{AB}\; \omega^A\;\omega^B\;\;\;,\;\;\; \eta = diag(-1,1,\dots,1)
\label{metrica11}
\ee
The  (pseudo) Riemannian connection is a one-form valued (in general) in the
$spin(1,10)$ algebra
\ba
\Omega &\equiv& \frac{1}{2}\; \omega^{AB}\;X_{AB}\cr
[X_{AB}\,;X_{CD}] &=& \eta_{AD}\; X_{BC} + \eta_{BC}\; X_{AD} - (A\leftrightarrow B)
\label{conexion}
\ea
and its components are fixed by the conditions of metricity and no-torsion
\be
\omega^{AB} = - \omega^{BA}\;\;\;,\;\;\;d\omega^A + \omega^{AB}\wedge\omega_B=0
\label{metritorsion}
\ee
For the metric (\ref{metrica11}) we get the following non-zero components
\ba
\omega_{0i} &=& \frac{1}{2}\;V^{-1}\;U^{-\frac{1}{2}}\;\left(
V^{\frac{1}{2}}\;U^{-\frac{1}{2}}\;\partial_i U\;(\omega^0 + \omega^5)
+ \epsilon_{ijk}\; B_k\;\omega^j-B_i\;\omega^4\right)= - \omega_{i5}\cr
\omega_{04} &=&-\frac{1}{2}\;V^{-1}\;U^{-\frac{1}{2}}\;B_i\;\omega^i = -\omega_{45}\cr
\omega_{05} &=&\frac{1}{2}\;V^{-\frac{1}{2}}\; U^{-1}\;\partial_i U\;\omega^i =
d\ln U^\frac{1}{2}\cr
\omega_{i4} &=&
\frac{1}{2}\;V^{-\frac{3}{2}}\;\left(
V^{\frac{1}{2}}\;U^{-\frac{1}{2}}\; B_i\;(\omega^0 + \omega^5)
+ \epsilon_{ijk}\;\partial_j V\;\omega^k + \partial_i
V\;\omega^4\right)
= -\frac 12 \epsilon_{ijk}\omega_{jk}
\label{conexion11}
\ea
where $\; B_i=\epsilon_{ijk}\,\partial_j\,A_k\;$ and indices are raised and
lowered with $(\eta_{AB})$.

The curvature tensor is a two-form valued (in general) in $spin(1,10)$
\ba
R &\equiv& d\Omega + \Omega\wedge \Omega = \frac{1}{2}\;R^{AB}\;X_{AB}\cr
R_{AB} &\equiv& d\omega_{AB} + \omega_{AC}\wedge\omega^C{}_B = \frac{1}{2}\; R_{ABCD}\;
\omega^C\wedge\omega^D\label{curvatura}
\ea
It is straightforwardly computed from (\ref{elfbein11}), (\ref{conexion11}); we get\footnote{
It is worth to note a subtlety that arises when ``magnetic" sources coupled to gravity are
present.
While the properties of the curvature tensor $R_{ABCD}= -R_{ABDC} = -R_{BACD}$ follow from
the two-form condition and metricity respectively more or less by definition, the torsionless
condition in (\ref{metritorsion}) leads to the anomaly relation
\be
R_{ABCD} - R_{CDAB} = -\frac{1}{2}\; (d_{ABCD} - d_{CDAB} + d_{DABC} - d_{BCDA})
\label{anomalia}
\ee
where we have defined
\ba
d^2 \omega^A &\equiv& \frac{1}{3!}\,d^A{}_{BCD}\;\omega^B\wedge\omega^C\wedge\omega^C\cr
d^A{}_{BCD} &\equiv& e_B{}^M\; e_C{}^N\; e_D{}^P\; \left(
[\partial_M\,;\partial_N]\omega_P{}^A+ [\partial_P\,;\partial_M]\omega_N{}^A
+[\partial_N\,;\partial_P]\omega_M{}^A\right)\label{anomaliacoef}
\ea
In ``normal" conditions (when the vielbein is smooth enough such that derivatives commute
on them) the coefficients $\{d_{ABCD}\}$ are identically null and the textbook property
$R_{ABCD}=R_{CDAB}$ holds; however when "Dirac string" like singularities are present its generalization
(\ref{anomalia}) must be considered.
In the case at hand (\ref{elfbein11}), (\ref{metrica11}) it can be shown by using
(\ref{constraintsbis}) that the non-trivial anomaly terms are
\ba
R_{i4jk}-R_{jki4} &=& \frac{1}{2}\; d_{4ijk} = \frac{1}{2}\;\epsilon_{ijk}\;
V^{-2}\,\nabla^2 V\cr
R_{04ij}-R_{ij04} &=& \frac{1}{2}\; d_{40ij} = \frac{1}{2}\; \epsilon_{ijk}\;
U^{-\frac{1}{2}}\; V^{-\frac{3}{2}}\,\nabla^2 A_k =
R_{54ij}-R_{ij54} = \frac{1}{2}\; d_{45ij} =
R_{4i0j}-R_{0j4i}= R_{4i5j}-R_{5j4i}\cr
& &
\ea
Explicit computation shows them up in (\ref{curvatura11}).
}
\ba
R_{0i} &=& \frac{1}{4}\;U^{-1}\;V^{-2}\;\left(\left( 4\;B_i\;B_j + \delta_{ij}\;
(\vec\nabla U\cdot\vec\nabla V - \vec B{}^2 )+ 2\;V\;\partial_i\partial_j U
\right.\right.\cr
&&- \left.\left. \partial_i U\;\partial_j V\; -
\partial_j U\;\partial_iV\right)\; \omega^j
-\epsilon_{ijk}\;\partial_j U\;\partial_k V\;\omega^4 \right)\wedge (\omega^0 + \omega^5)\cr
&&+ \frac{1}{2}\, U^{-\frac{1}{2}}\;V^{-\frac{5}{2}}\; \left(
\vec B\cdot\vec\nabla V\;\delta_{il} - 2\;B_i\;\partial_l V - B_l\;\partial_i V
+ V\;\partial_i B_l \right) \omega^4\wedge\omega^l\cr
&&+
\frac{1}{2}\,\epsilon_{ljk}\left(R_{0i4l}+\frac 12 U^{-\frac 12}V^{-\frac 32}\epsilon_{lim}\nabla^2A_m\right)\omega^j\wedge\omega^k    \cr
R_{04} &=& \frac{1}{4}\;U^{-1}\;V^{-2}\; \left(-(\vec B{}^2 + \vec\nabla U \cdot\vec\nabla V)
\;\omega^4 - \epsilon_{ijk}\;\partial_j U\;\partial_k\; V\;\omega^i\right)\wedge
(\omega^0 + \omega^5)\cr
&&- \frac{1}{2}\; \epsilon_{ijk}\;\left( R_{k404} -
\frac{1}{2}\; U^{-\frac{1}{2}}\; V^{-\frac{3}{2}}\;\nabla^2 A_k \right)\;\omega^i\wedge\omega^j
\cr
R_{05} &=& 0\cr
R_{i4} &=&  \left(\left(-R_{0ji4} + \frac{1}{2}\; U^{-\frac{1}{2}}\; V^{-\frac{3}{2}}\;\epsilon_{ijk}\nabla^2 A_k
\right)
\;\omega^j -\frac{1}{2}\; U^{-\frac{1}{2}}\; V^{-\frac{5}{2}}\;\epsilon_{ijk}\; B_j\;\partial_k V
\;\omega^4\right)\wedge (\omega^0 + \omega^5)\cr
&+& \frac{1}{2}\; V^{-3}\; \left( \vec\nabla V\cdot\vec\nabla V\;\delta_{ij}
-3\;\partial_i V\;\partial_j V + V\;\partial_j \partial_i V \right)
\omega^j\wedge\omega^4\cr
&-&\frac{1}{2}\;\epsilon_{ljk}\; (R_{l4i4}-\frac{1}{2}\; V^{-2}\;\nabla^2 V\;\delta_{li})\;
\omega^j\wedge\omega^k
\label{curvatura11}
\ea
while the other components are obtained from the relations
\be
R_{i5} = -R_{0i}\;\;\;,\;\;\; R_{45} = -R_{04}\;\;\;;\;\;\; R_{ij} = -\epsilon_{ijk}\; R_{k4}
\label{curvatura11bis}
\ee
The non-zero components of the Ricci tensor
$\; {\cal R}_{AB}\equiv R^C{}_{ACB}={\cal R}_{BA}\;$ and Ricci scalar result
\ba
{\cal R}_{00} &=& -\frac{1}{2}\; V^{-1}\;U^{-1}\;\nabla^2 U = {\cal
R}_{55}\;\;\;,\;\;\;{\cal R}_{ij} = - {\cal R}_{44}\;\delta_{ij}\;\;\;,\;\;\;
{\cal R}_{44} = \frac{1}{2}\; V^{-2}\;\nabla^2 V\cr
&&\;\;\;\;\;\;\;\;\;\;\;\;\;\;\;\;\;\;\;\;\;\;\;\;\;\;\;{\cal R} \equiv {\cal R}^A{}_A = -V^{-2}\;\nabla^2 V
\ea
This shows that the metric
is a solution of  $D=11$ SUGRA coupled to an object
living on the curve ${\cal C}$, being Ricci flat outside it\footnote{
Let us remind the reader that from (\ref{funcionesdelocalizadas}) it follows that
\be
-\nabla^2\left\{\begin{array}{l}U(\vec y)\cr V(\vec y)\cr\vec A (\vec
y)\end{array}\right. = \int_{\cal C} d\sigma\;\delta^3(\vec y - \vec y (\sigma))\;
\left\{\begin{array}{l}|\pi(\sigma)|\cr|B(\sigma)|\cr N\, {\vec
y}'(\sigma)\end{array}\right. \ee }.

\subsection*{Holonomy group and Killing spinors}

It is clear that, being reducible, the holonomy group of (\ref{metrica15}) will be
$\; Hol(ds_{11}^2) = Hol(ds_{1,5}^2)\times Hol(d\vec x\cdot d\vec x) =
Hol(ds_{1,5}^2)\subset Spin(1,5)\;$, but being the solution $\frac{1}{4}$-supersymmetric
(see below for more) the holonomy will be further reduced. To see this let us introduce
the light-cone generators $\;X^{\pm a}\equiv X^{0a}\pm X^{5a}\;\;,\;a=1,2,3,4\;\;,\;\;
X^{+-} = -2\,X^{05}$ and in terms of them the following (redundant) ones
\ba
T_i^{(\pm)}&=& \frac{1}{4}\;\epsilon_{ijk}\;X_{jk} \pm \frac{1}{2}\;X_{i4}\;\;\;,\;\;\;i=1,2,3\cr
t_\pm^1 &=& X^{+3}\mp i\, X^{+4}\;\;\;,\;\;\; \bar t_\pm^1 = X^{+3}\pm i\, X^{+4}\cr
t_\pm^2 &=& X^{+1} +i\, X^{+2}\;\;\;,\;\;\;\bar t_\pm^2 = X^{+1}- i\, X^{+2}
\ea
leaving aside $\;\{ X^{-a},\; a=1,2,3,4,+ \}\;$ which will not play any role in the discussion.
It is easy to check the following commutation relations
\ba
[T_i^{(\pm)}\,;T_j^{(\pm)}]&=&
-\epsilon_{ijk}\;T_k^{(\pm)}\;\;\;,\;\;\; [T_i^{(+)}\,;T_j^{(-)}]= 0\cr [T_i^{(\pm)}\,;
t_\pm^{\alpha}]&=& -\frac{i}{2}\; \sigma_i{}^\alpha{}_\beta\; t_\pm^{\beta}
\;\;\;,\;\;\; [T_i^{(\pm)}\,; \bar t_\pm^{\alpha}]= \frac{i}{2}\;
\sigma_i^*{}^\alpha{}_\beta\; \bar t_\pm^{\beta}\cr
[t_\pm^{\alpha}{}\,;t_\pm^{\beta}]&=& [t_\pm^{\alpha}{}\,;\bar t_\pm^{\beta}] =
[t_+^{\alpha}{}\,;t_-^{\beta}] = [t_+^{\alpha}{}\,;\bar t_-^{\beta}] =0\label{subalgebra}
\ea
The first line is just the
$spin(4) \sim su(2)_+\times su(2)_-$ subalgebra of $spin(1,5)$, while the second line
indicates that $\; (t_\pm^{\alpha})\;$ and $\; (\bar t_\pm^{\alpha})\;$ transform in the
spin $\frac{1}{2}$ representation of $su(2)_\pm$ and its conjugate (equivalent anyway)
respectively.
Finally the third line implies that $\;\{t_\pm^{\alpha}\;,\;\bar t_\pm^{\alpha}\}\;$
expand an abelian subalgebra.

Now from the curvature tensor we can read the Lie algebra $hol(ds_{11}^2)$ of
$Hol(ds_{11}^2)$; from  (\ref{curvatura11}),  (\ref{curvatura11bis}) it takes  the form
\be
R = R_{0i}\; T_i^{(-)} + \frac{1}{2}\;\left( (R_{03}-i\,R_{04})\;t_-^1 +
(R_{01}-i\,R_{02})\;t_-^2 + h.c. \right)
\ee
which shows that $\;hol(ds_{11}^2)\;$ is the seven dimensional subalgebra whose elements can be written as
\be
X = \epsilon_i\;T_i^{(-)}- + \epsilon_\alpha\;  t_-^\alpha + \bar\epsilon_\alpha\;\bar t_-^\alpha
\;\;\;,\;\;\;\epsilon_i \in\br\;\;\;,\;\;\;  \bar\epsilon_\alpha =\epsilon_\alpha^*
\label{section}
\ee
It is given by
\ba
&&[T_i^{(-)}\,;T_j^{(-)}]=
-\epsilon_{ijk}\;T_k^{(-)}
\;\;\;,\;\;\;
[T_i^{(-)}\,;t_{-}^{\alpha}]= -\frac{i}{2}\; \sigma_i{}^\alpha{}_\beta\; t_{-}^{\beta}
\;\;\;,\;\;\;
[T_i^{(-)}\,; \bar t_{-}^{\alpha}]= \frac{i}{2}\;\sigma_i^*{}^\alpha{}_\beta\; \bar t_{-}^{\beta}
\cr
&&\;\;\;\;\;\;\;\;\;\;\;\;\;\;\;\;\;\;\;\;\;\;\;\;\;\;\;\;\;
[t_{-}^{\alpha}{}\,;t_{-}^{\beta}]= [t_{-}^{\alpha}{}\,;\bar t_{-}^{\beta}] =
[\bar t_-^{\alpha}{}\,;\bar t_-^{\beta}] = 0\label{subalgebraH}
\ea

On the other hand, in view of the pure geometric character of the solution we have that the
variation of the gravitino field $\;\chi^\Lambda_A\;$ under a SUGRA transformation is just
\be
\delta_\epsilon\,\chi^\Lambda_A = D_A\epsilon^\Lambda = e_A(\epsilon^\Lambda) +
S(\Omega)^\Lambda{}_{\Lambda'}\;
\epsilon^{\Lambda'}
\ee
where the parameter $\epsilon^\Lambda$ is a thirty two component Majorana spinor
 and $S(\Omega)$ is the connection (\ref{conexion}) valued in the
spinorial representation.
The solution will preserve the SUSY's that annihilate the gravitino variation, so the Killing
spinors will be the covariantly constant ones.
Specifically in a Weyl basis where for definiteness we take
\be
\Gamma_m = \left(\matrix{0_{16}&\gamma_m\cr \gamma_m &0_{16}  }\right)\;\;\;;\;\;\;
\Gamma_9 = i\,\left(\matrix{0_{16}& 1_{16}\cr -1_{16} &0_{16}  }\right)\;\;\;;\;\;\;
\Gamma_{10} = \left(\matrix{1_{16}& 0_{16}\cr 0_{16} & -1_{16}  }\right)
\ee
where $\;\{\gamma_m\;,\; m=0,1,\dots,8\}\;$ are gamma-matrices of $spin(1,8)$,
the condition $\;\delta_\epsilon\,\chi^\Lambda_A = D_A\epsilon^\Lambda = 0\;$
results equivalent to
\ba
\epsilon &=& U^{-\frac{1}{4}}\; \epsilon_0
\;\;\;,\;\;\;\mbox{with}\;\;\;
\left\{
\begin{array}{rlr}
\Gamma_{05}\;\epsilon_0 &=& -\epsilon_0\\ \Gamma_{1234}\;\epsilon_0 &=& -\epsilon_0
\end{array}
\right. \label{killing}
\ea
with $\epsilon_0$ a constant spinor.
Equation (\ref{killing}) signals the preservation of $8=\frac{1}{4}\;32$ of the
supersymmetries as it happened for the supertube.
\footnote{
We note that the Killing spinors can be made constant through a Lorentz boost
that eliminates the part of the connection that does
not lie in the holonomy algebra, given by $\;\omega_{05}\;S(X^{05})=d\ln
U^\frac{1}{4}\; \Gamma^{05}\;$ (see (\ref{conexion11})).
}
We can go a little bit further; in the spinorial representation the generators of
$hol(G^{(11)})$ take the form
\ba
S_1^{(-)} &=& \frac{1}{4}\;\Gamma_{13}\;(\Gamma_{12} - \Gamma_{34})\;\;\;,\;\;\;
S_2^{(-)} = \frac{1}{4}\;\Gamma_{23}\;(\Gamma_{12} - \Gamma_{34})\;\;\;,\;\;\;
S_3^{(-)} = \frac{1}{4}\;(\Gamma_{12} - \Gamma_{34})\cr
S_-^1 &=& -\frac{1}{2}\; (\Gamma_{35} + i\,\Gamma_{45})\,(1 + \Gamma_{05} )\;\;\;,\;\;\;
\overline S _-^1 = -\frac{1}{2}\; (\Gamma_{35} - i\,\Gamma_{45})\,(1 + \Gamma_{05} )\cr
S_-^2 &=& -\frac{1}{2}\; (\Gamma_{15} + i\,\Gamma_{25})\,(1 + \Gamma_{05} )\;\;\;,\;\;\;
\overline S _-^2 = -\frac{1}{2}\; (\Gamma_{15} - i\,\Gamma_{25})\,(1 + \Gamma_{05} )
\ea
From here it is clear that first and second conditions in (\ref{killing}) are
equivalent to ask for the trivial action of the abelian and $su(2)$ subalgebras
respectively on the space of spinors; we conclude that there exist exactly eight singlets
under the full holonomy algebra.
If w.r.t. the Cartan subalgebra
$\;\{\Gamma_{05}, i\,\Gamma_{12}, i\,\Gamma_{34}, i\,\Gamma_{67}, i\,\Gamma_{89}\}\;$
we label a basis by $(s_0 , s_1,\dots,s_4)$, $\,s_i=0,1\,$ (see i.e. Appendix of
\cite{lugo} for details), they are simply given by $(0 s_2 s_2 s_3 s_4)$.

\section{Discussion}

First let us note that the supersymmetry of the supergravity supertube solution does not depend
on any relation between the $D0$ and $F1$ charge densities, what enables us to assign
arbitrary values to them.

While the original supertube had zero $D2$ charge, the final
ten dimensional configuration (\ref{final}) is a genuine $D6$-brane, its
charge being traced to the initial $D0$ brane density,
\be
q_{D6}\equiv
\int_{S^2} *_{\it 10} F_{\it 8} =
\int_{S^2} F_{\it 2}=
\int_{S^2} *_{\it 3}dV=
\int_{
B_3, \;
\partial B_3=S^2
}
\!\!\!\!\!\!\!\!\!\!\!\! d*_{\it 3}dV = -\int_{{\cal C}} d\sigma |B(\sigma)|
\ee
Of course  for $U=1, \vec A = \vec 0$ (which means $|\pi(\sigma)|=0$ on a flat $D2$-brane) we recover the standard
flat $\frac 12$-supersymmetric $D6$-brane solution.

When uplifted to eleven dimensions, the resulting configuration (\ref{metrica15}) can be interpreted as a bound
state of a Taub-NUT like space and pp-wave. The presence of the first can be verified by direct inspection of
the metric, while that of the second follows from the existence of the covariantly constant null vector
$\partial_-$. This assertion is consistent with the analysis of the Killing spinors since the constraints
(\ref{killing}) are those corresponding to a Taub-NUT and a pp-wave supersymmetry preserving conditions
\cite{Gauntlett:1997cv}.

For $U=1, \vec A = \vec 0$ we have the product of Minkowski space in $1+6$ dimensions
and a Taub-NUT space, i.e. M-theory compactified on this last one which is known to
preserve $\frac 12$ of the supersymmetries and to have $SU(2)\subset G_2$ holonomy.
Since in our general $\frac 14$-supersymmetric case the irreducible part of the metric is
six dimensional one could expect to find a $Hol(ds^2_{1,5}) \subset SU(3)$ reduced
holonomy group, but the pseudo-Riemannian character of $ds^2_{1,5}$ invalidates Berger's
classification theorem \cite{Joyce:2001xt} and the resulting manifold is not a Calabi-Yau one\footnote{
At the time of writing we have no knowledge about a classification of holonomy groups of
pseudo-riemannian manifolds.
}. In fact, from (\ref{subalgebraH}) the holonomy group is non semisimple and locally isomorphic
to the semidirect product of $SU(2)$ and an the real section defined in (\ref{section}) of the Abelian
four-dimensional algebra generated by $\{t_-^\alpha,\bar t_-^\alpha \}$ transforming in the $\frac 12 \oplus \bar {\frac 12}$.

\section{Acknowledgments}

We would like to thank Carlos N\'{u}\~{n}ez for an enlightening discussion.

\end{document}